\documentclass[sigconf]{acmart}
\usepackage{subcaption}

\AtBeginDocument{%
  }

\newcommand{\cmark}{\ding{51}}
\newcommand{\xmark}{\ding{55}}

\copyrightyear{2026}
\acmYear{2026}
\setcopyright{cc}
\setcctype{by-nc-nd}
\acmConference[UbiComp Companion '26]{Companion of the 2026 ACM International Joint Conference on Pervasive and Ubiquitous Computing}{October 11--15, 2026}{Shanghai, China}
\acmBooktitle{Companion of the 2026 ACM International Joint Conference on Pervasive and Ubiquitous Computing (UbiComp Companion '26), October 11--15, 2026, Shanghai, China}
\acmDOI{10.1145/3798063.3842643}
\acmISBN{979-8-4007-2533-3/2026/10}

\begin{document}
\title{Understanding Behavioral Dark Patterns of High BMI Individuals}

\author{Manjeet Yadav}
\affiliation{%
    \department{Computer Science and Engineering}
	\institution{Indian Institute of Technology Jodhpur}
    \city{Jodhpur}
    \state{Rajasthan}
	\country{India}
}
\email{p25cs0011@iitj.ac.in}

\author{Prasenjit Karmakar}
\affiliation{%
    \department{Computer Science and Engineering}
	\institution{Indian Institute of Technology Kharagpur}
    \city{Kharagpur}
    \state{West Bengal}
	\country{India}
}
\email{prasenjitkarmakar52282@gmail.com}

\author{Suchetana Chakraborty}
\affiliation{%
    \department{Computer Science and Engineering}
	\institution{Indian Institute of Technology Jodhpur}
    \city{Jodhpur}
    \state{Rajasthan}
	\country{India}
}
\email{suchetana@iitj.ac.in}

\begin{abstract}
Understanding how everyday behaviors influence body weight is essential for designing effective and personalized health interventions. Existing studies largely rely on self-reported questionnaires or limited sensing modalities, making it difficult to capture the temporal dynamics of daily behavior. In this work, we analyze the \textit{DiversityOne} dataset, comprising four weeks of passive smartphone sensing and ecological momentary assessments collected from $453$ university students across eight countries. We extract behavioral features spanning dietary habits, physical activity, screen time, and smartphone usage, and investigate their associations with self-reported Body Mass Index (BMI). Beyond feature-level analysis, we employ Hidden Markov Models (HMMs) to uncover latent behavioral patterns. Our analysis reveals that higher BMI is associated with more frequent consumption of soda, alcohol, and processed meat. We further reveal that overweight and obese individuals spend longer periods in food delivery apps and are more likely to transition back to unhealthy eating and drinking routines after starting to exercise. In contrast, normal-weight individuals lead a more balanced lifestyle. These findings highlight key behavioral patterns that make weight loss particularly challenging.
\end{abstract}



\begin{teaserfigure}
  \centering 
  \includegraphics[width=0.9\textwidth]{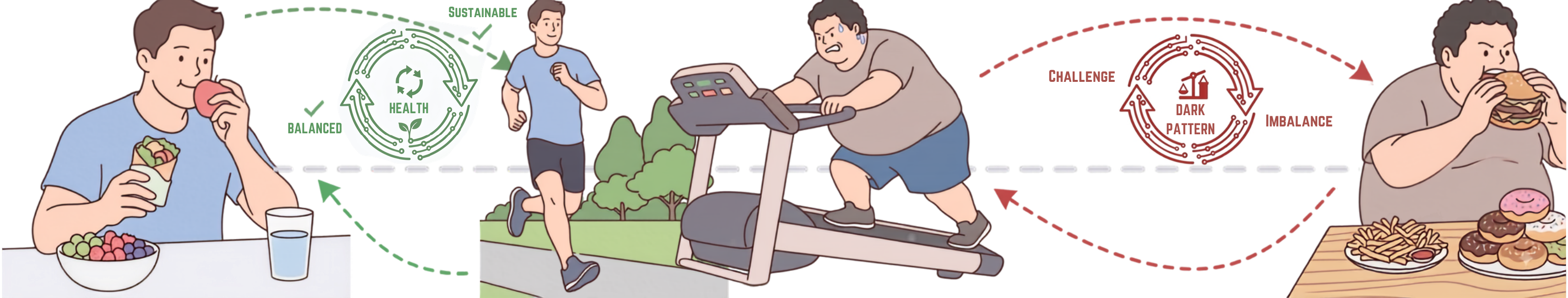}
  \caption{Our work aims to identify and understand contrasting behavioral cycles associated with body weight using passive smartphone sensing. Healthy individuals typically maintain a sustainable loop of balanced eating and moderate physical activity that reinforces healthy habits. In contrast, high-BMI individuals often exhibit a behavioral \textit{dark pattern}, where bouts of excessive exercise are followed by relapse into processed foods and sugar-sweetened beverages, creating a self-reinforcing vicious cycle that hinders long-term weight management.}
  \Description{Our work aims to identify and understand contrasting behavioral cycles associated with body weight using passive smartphone sensing. Healthy individuals typically maintain a sustainable loop of balanced eating and moderate physical activity that reinforces healthy habits. In contrast, high-BMI individuals often exhibit a behavioral \textit{dark pattern}, where bouts of excessive exercise are followed by relapse into processed foods and sugar-sweetened beverages, creating a self-reinforcing vicious cycle that hinders long-term weight management.}
  \label{fig:teaser_obese}
\end{teaserfigure}

\maketitle

\section{Introduction}

Overweight and obesity have become major public health challenges, affecting more than $2.5$ billion adults worldwide and increasing the risk of long-term health complications such as cardiovascular diseases, diabetes, cancers, neurological disorders, chronic respiratory diseases, and digestive disorders~\cite{WHO2025Obesity}. Obesity is driven not by any single behavior but by an interlocking system of dietary composition, physical inactivity, food environments, and increasingly digital engagement~\cite{cong2025digital,younossi2023lifestyle,zink2024longitudinal}. Existing studies~\cite{wang2026uncovering,bycroft2018uk,fain2017nhanes,liu2018network,xu2023globem} rely heavily on self-reported questionnaires and food diaries, which are expensive to administer, prone to recall bias, and provide only sparse snapshots of everyday behavior. Consequently, they fail to capture the temporal behavioral patterns that ultimately contribute to weight gain. In contrast, smartphones offer a scalable and unobtrusive platform for studying everyday lifestyle behaviors. Equipped with a rich array of built-in sensors, they continuously capture information about physical activity, mobility, screen usage, and application interactions~\cite{chen2026beyond}. In addition, ecological momentary assessments (EMA), delivered through notifications and pop-ups, enable users to report meals, snacks, and other contextual information in situ with minimal recall bias~\cite{gross2023experience}.

\begin{table}
\centering
\scriptsize
\caption{Smartphone datasets to realize behavioral dynamics.}
\label{tab:datasets}
\begin{tabular}{lccccccl} 
\toprule
\textbf{Dataset}                                           & \textbf{\# User} & \textbf{\# Country} & \textbf{Screen}   & \textbf{Apps} & \textbf{Diet} & \textbf{Activity}   &\textbf{BMI} 
\\ 
\midrule
DiversityOne~\cite{busso2025diversityone} & 782                   & 8                   & \cmark                & \cmark            & \cmark            & \cmark                  &\cmark           
\\
StudentLife~\cite{wang2014studentlife}     & 48                    & 1                   & \cmark                & $\sim$            & \xmark            & \cmark                  &\xmark           
\\
LifeSnaps~\cite{yfantidou2022lifesnaps}   & 71                    & 4~                  & \xmark                & \xmark            & \xmark            & \cmark                  &\xmark           
\\
GLOBEM~\cite{xu2023globem}                & 497                   & 1                   & \xmark                & $\sim$             & \xmark            & \cmark                  &\cmark           
\\
NetHealth~\cite{liu2018network}           & 700                   & 1                   & \xmark                & $\sim$             & \xmark            & \cmark                  &\cmark           
\\
NHANES~\cite{fain2017nhanes}               & 10K                   & 1                   & $\sim$ & \xmark            & $\sim$     & \xmark                  &\cmark           
\\
UK Biobank~\cite{bycroft2018uk}           & 502K                  & 1                   & $\sim$   & \xmark            & $\sim$     & \cmark                  &\cmark           \\
\bottomrule
\end{tabular}
\end{table}

Despite the growing use of smartphone sensing to characterize everyday behaviors~\cite{bangamuarachchi2025inferring,haucke2024social,nepal2024moodcapture,assi2023complex}, its relationship with body weight remains underexplored. Prior studies have predominantly relied on cohorts from a single country~\cite{wang2014studentlife,campana2021contextlabeler,laporte2023laureate,xu2023globem}, and are prone to cultural and environmental bias (See \tablename~\ref{tab:datasets}). Many datasets are drawn from self-selected users of health and weight-loss applications~\cite{veluvali2025impact}, thereby limiting the generalizability of their findings to broader populations. Furthermore, self-reported dietary information is known to be systematically biased, particularly among individuals with higher BMI~\cite{veluvali2025impact}. However, its implications for smartphone-based behavioral studies remain largely unexplored.

Recently released \textit{DiversityOne} dataset ~\cite{busso2025diversityone} provides a unique opportunity to address these limitations. The dataset contains approximately four weeks of passive smartphone sensing and intensive EMA data collected from $453$ University students (Participants who reported BMI category and contributed data for at least one week) across eight countries spanning four continents. In addition to passive sensing streams, participants reported meals, snacks, activities, and other contextual information, which makes \textit{DiversityOne} well-suited for investigating behavioral correlations of BMI in everyday life. Our key research questions are:
\begin{itemize}
    \item \textbf{RQ1:} Which lifestyle choices (i.e., diet, physical activity, digital engagement, etc.) show consistent association with BMI across cultures?
    \item \textbf{RQ2:} How does the daily behavioral routine of obese individuals make weight loss challenging?
\end{itemize}

To answer these questions, we have conducted Spearman rank correlation tests and modeled the temporal behavioral data using a Hidden Markov Model (HMM). Our analysis reveals that greater consumption of alcohol $(\rho = 0.17, q < 0.001)$, soda $(\rho = 0.16, q < 0.05)$, and processed meat $(\rho = 0.13, q < 0.05)$ is associated with higher BMI categories. For instance, alcohol consumption per meal increases from $3.5\%$ among normal individuals to $9.8\%$ among obese individuals, nearly a threefold increase. Moreover, obese people spend more time on food delivery apps $(\rho = 0.219, q < 0.001)$ and are likely to relapse back to junk eating after high-intensity exercise sessions. In contrast, normal-weight people exhibit a more sustainable lifestyle with occasional unhealthy escapes. This paper analyzes these daily behavioral dynamics in the \textit{DiversityOne} dataset to uncover hidden dark patterns associated with body weight.

\begin{figure}
    \centering
    \includegraphics[width=0.95\linewidth]{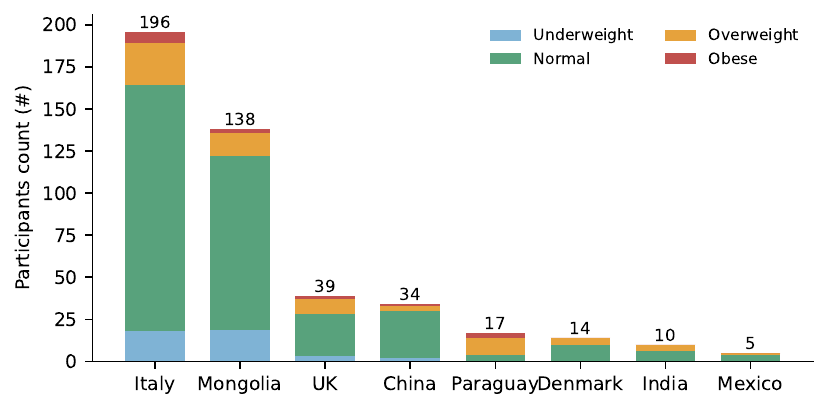}
    \caption{Number of participants by country and BMI distribution. Selected 453 participants include 42 underweight, 326 normal-weight, 70 overweight, and 15 obese individuals.}
    \Description{Number of participants by country and BMI distribution. Selected 453 participants include 42 underweight, 326 normal-weight, 70 overweight, and 15 obese individuals.}
    \label{fig:no_of_participants}
\end{figure}

\section{Dataset Preparation}
We use the \textit{DiversityOne} dataset~\cite{busso2025diversityone}, a large-scale mobile sensing dataset designed to study everyday behavior across multiple countries. The study recruited $782$ university students over four weeks from eight universities in Italy, Mongolia, the United Kingdom, China, Paraguay, Denmark, India, and Mexico. We selected $453$ participants who contributed at least one week of data and completed an exit questionnaire that included self-reported BMI categories: underweight (UW), normal (N), overweight (OW), or obese (OB). \figurename~\ref{fig:no_of_participants} shows the country-wise distribution of BMI categories. Participants installed the \textit{iLog} smartphone application, which continuously collected $26$ modalities, including physical activity, screen usage, application usage, and device status. Moreover, participants completed ecological momentary
assessments (EMA) at regular intervals throughout the day, reporting their current activity, location, social context, mood, and eating patterns. This unique combination of passive sensing, detailed dietary self-reports, and BMI data in \textit{DiversityOne} enables behavioral analysis of body weight.

\subsection{Data Pre-processing}
While analyzing the dataset, we identified several data-quality issues in the raw smartphone logs that could affect subsequent research findings if left unaddressed. The primary data inconsistencies and processing steps are discussed below.
\subsubsection{Duplicate Participant Identifiers:} Participant IDs were unique only within each country, resulting in identifier collisions across countries (182 IDs were reused). We therefore created globally unique identifiers by combining the country and participant ID.
\subsubsection{Step-counter Resets:} Android reports cumulative step counts, so device restarts and sensor resets produced implausible daily totals (exceeding 2 million steps per day in some cases). We estimated daily steps by summing only positive consecutive differences and capping the accumulation rate at five steps per second.
\subsubsection{Screen Sessions at Midnight:} Screen sessions crossing midnight were incorrectly assigned to a single calendar day. We recomputed complete \texttt{SCREEN\_ON} to \texttt{SCREEN\_OFF} sessions, split sessions at midnight, and allocated screen time to the appropriate day. Sessions longer than $180$ minutes were discarded as artifacts.
\subsubsection{Inflated Application Usage:} Foreground application logs are sampled periodically and may continue while the smartphone screen is off, causing raw polling counts to overestimate usage. We therefore estimated application usage only when the screen was on, rather than using raw durations.

\begin{figure*}
    \centering
    \begin{subfigure}{0.24\textwidth}
        \centering
        \includegraphics[width=\linewidth,keepaspectratio]{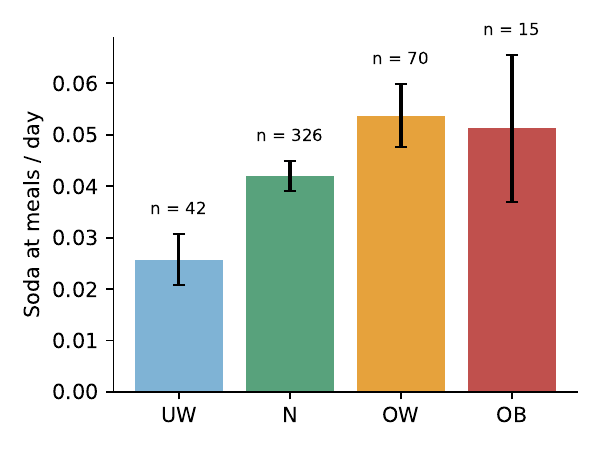}
        \caption{Soda Consumption}
        \label{fig:soda_consupmtion}
    \end{subfigure} 
    \begin{subfigure}{0.24\textwidth}
        \centering
        \includegraphics[width=\linewidth,keepaspectratio]{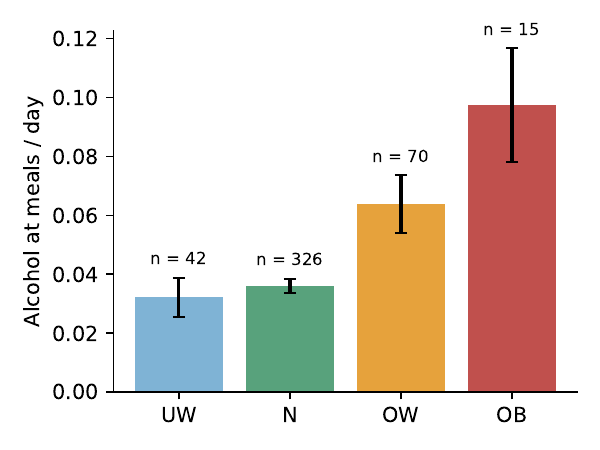}
        \caption{Alcohol Consumption}
        \label{fig:alcohol_consumption}
    \end{subfigure}
    \begin{subfigure}{0.24\textwidth}
        \centering
        \includegraphics[width=\linewidth,keepaspectratio]{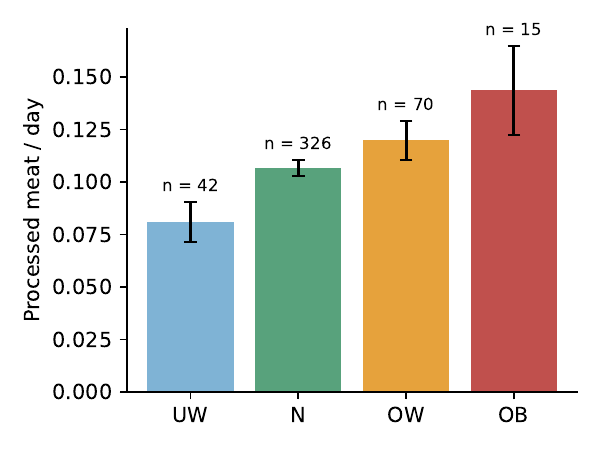}
        \caption{Meat Consumption}
        \label{fig:processed_meat}
    \end{subfigure}
    \begin{subfigure}{0.24\textwidth}
        \centering
        \includegraphics[width=\linewidth,keepaspectratio]{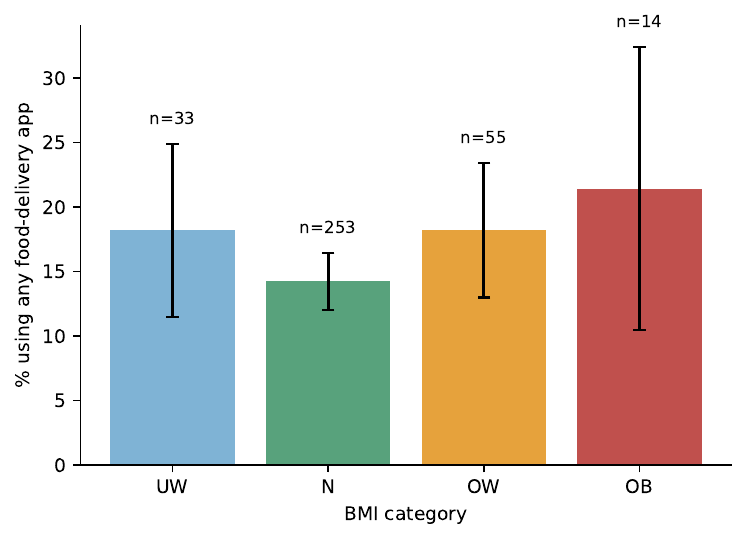}
        \caption{Food Delivery App Use}
        \label{fig:food_delivary_country_adjusted}
    \end{subfigure}
    \caption{Distribution of (a) soda consumption per day, (b) alcohol consumption per day, (c) processed meat consumption per day, and (d) percentage of participants using food delivery apps, against BMI categories.}
    \Description{Distribution of (a) soda consumption per day, (b) alcohol consumption per day, (c) processed meat consumption per day, and (d) percentage of participants using food delivery apps, against BMI categories.}
    \label{fig:bmi_assoication_behavior}
\end{figure*}

\subsection{Feature Extraction}
\label{sec:feat_extraction}

\begin{figure}
    \centering
    \includegraphics[width=0.95\linewidth]{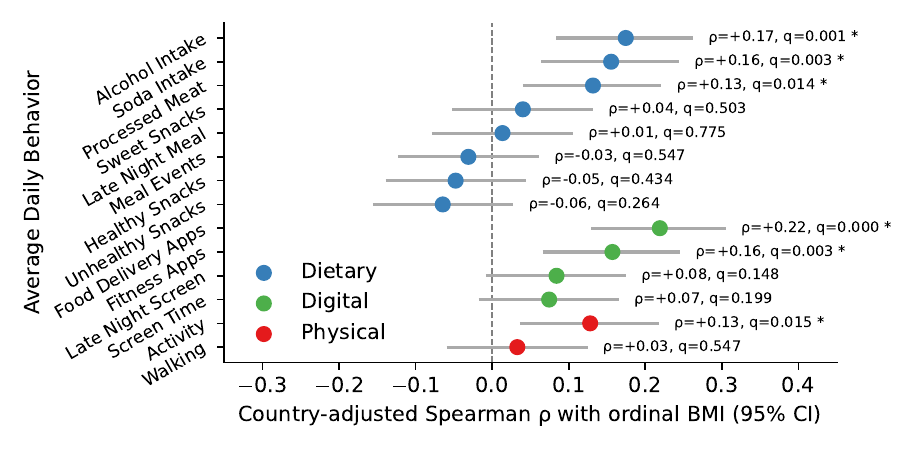}
    \caption{Associations between participants' average daily behavior and BMI categories across countries. The asterisk (*) represents statistically significant associations.}
    \Description{Associations between participants' average daily behavior and BMI categories across countries. The asterisk (*) represents statistically significant associations.}
    \label{fig:behaviour_variable_association}
\end{figure}

\figurename~\ref{fig:behaviour_variable_association} shows the Spearman rank correlation coefficients between participants' average daily behavioral features and their BMI categories. The features span the participants' dietary, digital, and physical activities. 
Dietary features are computed from the EMA, polled every 30 minutes, and include the daily counts of \textit{meal events}, \textit{late-night meals}, \textit{sweet snacks}, \textit{processed meat}, \textit{soda intake}, \textit{alcohol intake}, \textit{unhealthy}, and \textit{healthy} snack intake. Digital features included daily \textit{screen time}, \textit{late-night screen} use, and use of \textit{food delivery apps} and \textit{fitness apps} from the \textit{iLog} application logs. The physical \textit{activity} feature comprised daily counts of \textit{walking}, running, and bicycling, identified at 30-second intervals by the Google activity-recognition API~\cite{googleActivity} on the smartphone.

Due to varying sampling rates, we aggregated features across the day, yielding a sequence of participants' daily dietary, digital, and physical behaviors over the data collection period. Next, we discuss the observations from our analysis.

\begin{figure*}
    \centering
    \begin{subfigure}{0.41\linewidth}
        \includegraphics[width=\columnwidth,trim={0 0.8cm 0 0},clip]{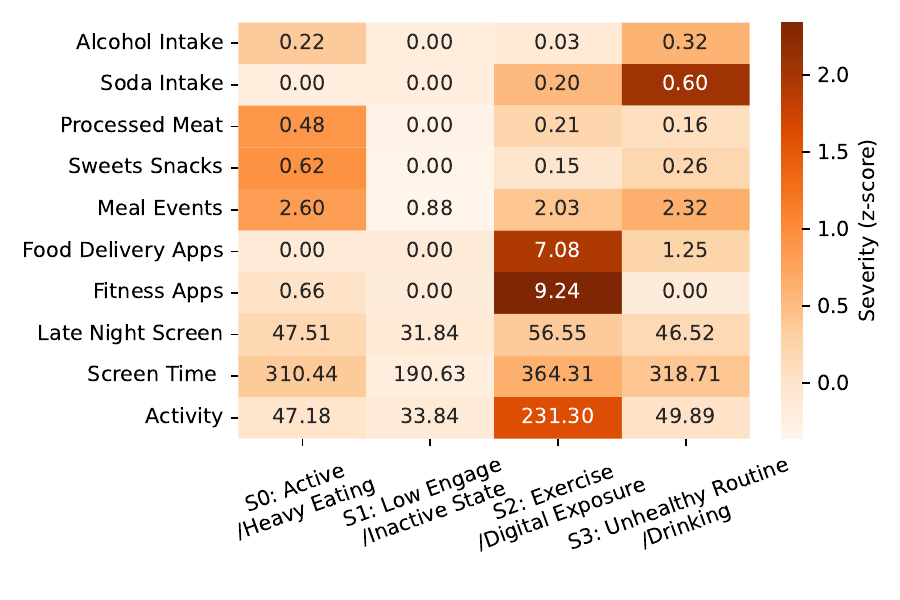}
        \caption{OB/OW latent states}
        \label{fig:HMM_result}
    \end{subfigure}
    \begin{subfigure}{0.25\linewidth}
        \centering
        \includegraphics[width=\columnwidth]{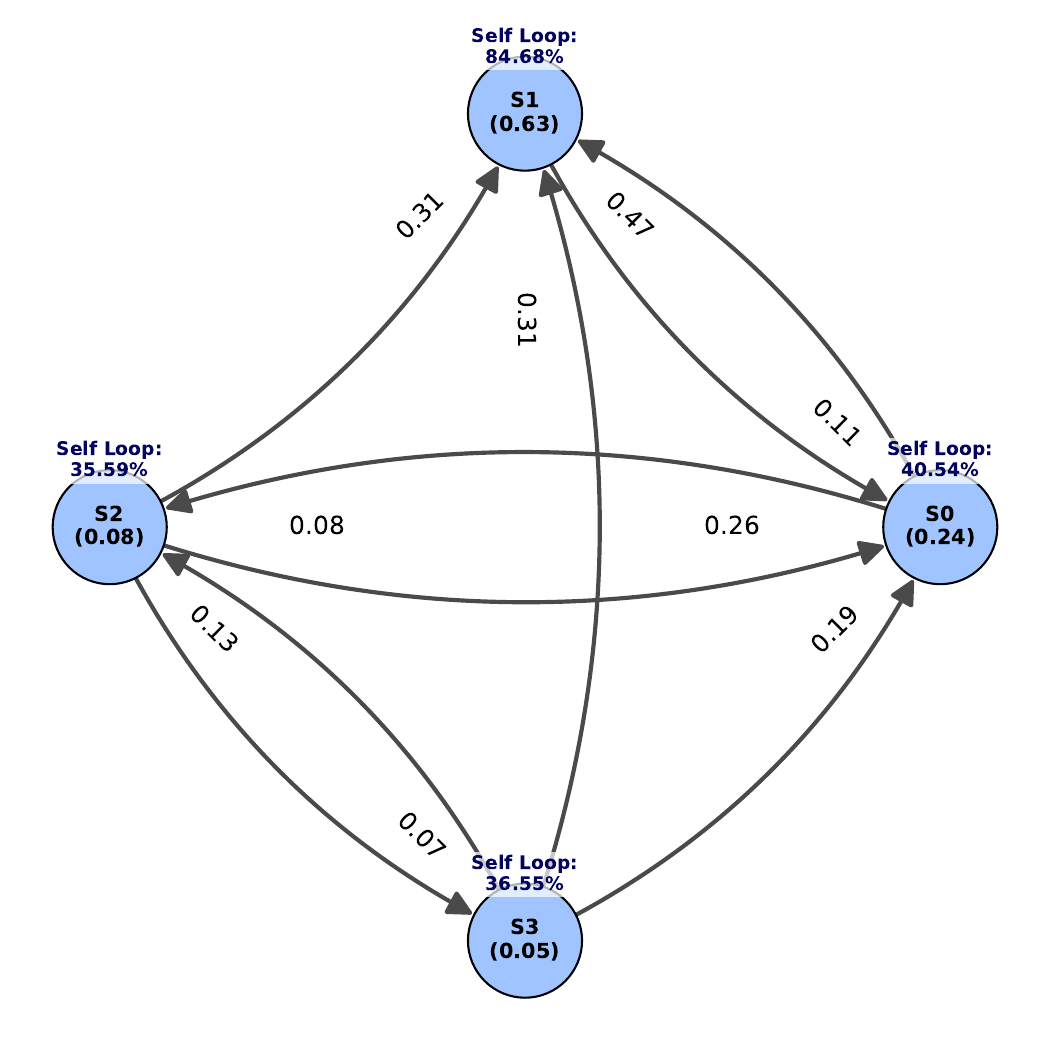}
        \caption{N/ UW Routine}
        \label{fig:normal_transition}
    \end{subfigure}
    \begin{subfigure}{0.25\linewidth}
        \centering
        \includegraphics[width=\columnwidth]{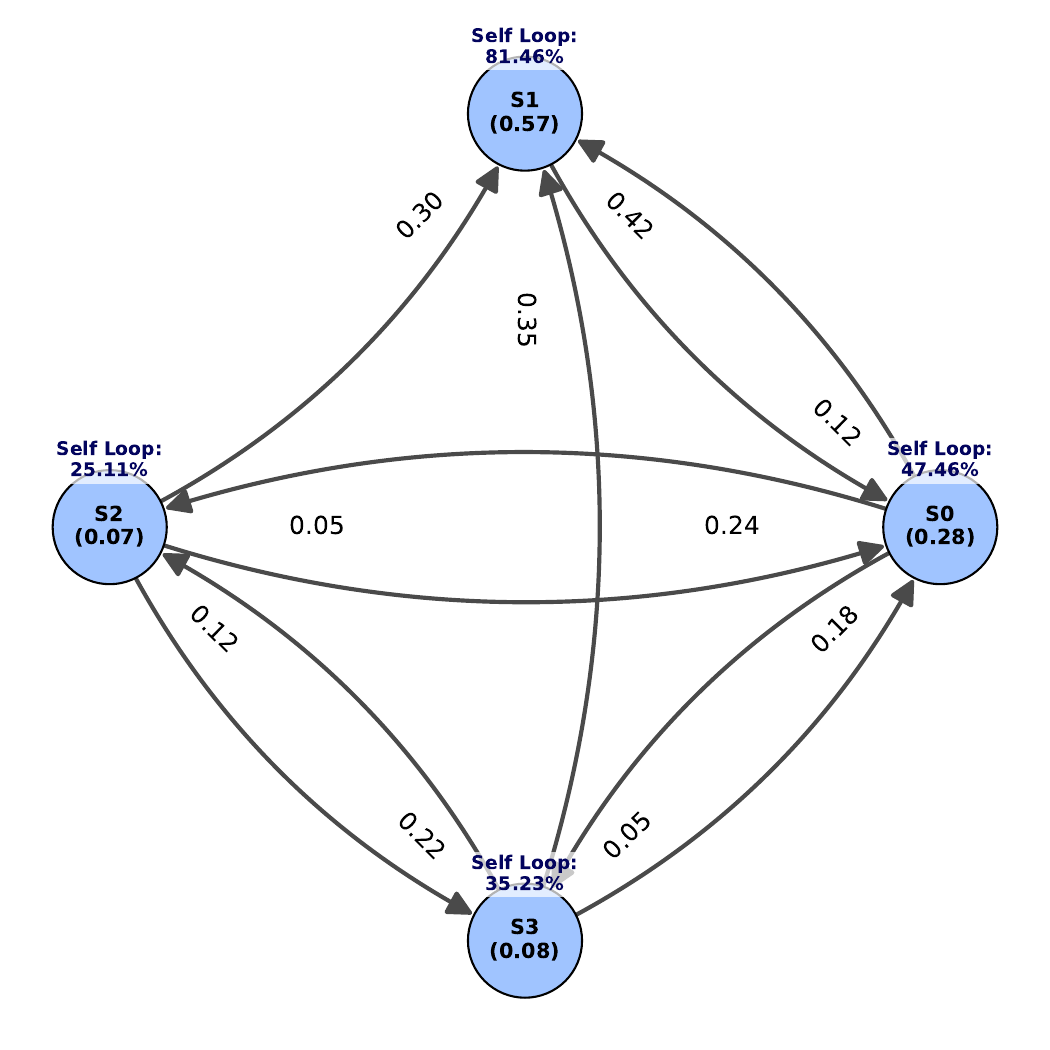}
        \caption{OB/ OW Routine}
        \label{fig:obesity_transition}
    \end{subfigure}
    \caption{Hidden Markov Model (a) latent behavioral states in obese and overweight users. Behavioral state transitions for (b) normal and underweight, and (c) obese and overweight individuals. For normal and underweight users, S0 represents healthy eating, and exercise intensity reduces for S2.}
    \Description{Hidden Markov Model (a) latent behavioral states in obese and overweight users. Behavioral state transitions for (b) normal and underweight, and (c) obese and overweight individuals. For normal and underweight users, S0 represents healthy eating, and exercise intensity reduces for S2.}
    \label{fig:HMM_analysis}
\end{figure*}

\section{Observations}
This section analyses participants' behaviors that show a consistent association with BMI across cultures (\textbf{RQ1}) and with daily routines that make weight loss challenging among obese individuals (\textbf{RQ2}).

\subsection{RQ1 -- Behavior vs BMI}
\label{sec:observation_rq1}
In this analysis, we used average daily behavioral features to examine the association between participants' consistent lifestyle choices and their BMI category. We observe that \textit{dietary behaviors} exhibited the strongest associations with BMI categories and are significantly positively correlated with BMI after FDR correction (see \figurename~\ref{fig:behaviour_variable_association}). Participants with higher BMI reported more frequent consumption of soda $(\rho = 0.16, q < 0.05)$, alcohol     $(\rho = 0.17, q < 0.001)$, and processed meat $(\rho = 0.13, q < 0.05)$ during meals. In contrast, sweets consumed during meals, meal timing, and snack frequency show positive associations with BMI but were not statistically significant. Healthy and unhealthy snack types show inconsistent associations with BMI (i.e., both are negatively correlated with BMI) in the dataset; we \textit{exclude} these features from our temporal analysis. Further, we \textit{exclude} late-night meals from our temporal analysis, as it shows very weak associations with BMI. \figurename~\ref{fig:bmi_assoication_behavior}a-c illustrates the dietary trends across BMI categories. For example, the proportion of meals containing alcohol increased from approximately $3.5\%$ among participants with normal BMI to $9.8\%$ among participants with obesity. Similar upward trends were observed for soda and processed meat consumption. These findings are consistent with prior epidemiological evidence~\cite{younossi2023lifestyle} linking alcohol consumption and processed meat intake to higher BMI, providing external validity for our findings.

Further, \figurename~\ref{fig:behaviour_variable_association} shows the strongest positive association between BMI and \textit{digital behavior} like usage of food-delivery apps $(\rho = 0.219, q < 0.001)$, and fitness apps $(\rho = 0.16, q < 0.001)$. As shown in \figurename~\ref{fig:food_delivary_country_adjusted}, we observe an increase in food-delivery apps usage from 14\% in normal-weight participants to 18\% in overweight and to 21\% in obese participants. In contrast, daily or late-night screen time shows statistically insignificant positive associations with BMI. These findings suggest that specific app usage may serve as a useful proxy for BMI.

Finally, \textit{physical behavior}, including daily activities (walking, running, cycling), shows a significant positive association with BMI. We \textit{exclude} the specific walking feature-subset from our temporal analysis. Overall, these observations suggest that dietary behaviors, food- or fitness-related digital behaviors, and physical activities explain variations in BMI among the participants in the dataset.

\subsection{RQ2 -- Routine vs BMI}
To characterize routine behavioral patterns across BMI groups, we used temporal dietary, digital, and physical behavioral features (i.e., excluding healthy and unhealthy snacks, late-night meals, and walking) based on average behavior vs. BMI analysis in Section~\ref{sec:observation_rq1}. We trained a Hidden Markov Model (HMM) on participants' daily behavioral sequences to identify differences in recurring latent behavioral states between individuals with normal and high BMI. Note that we normalized the feature sequences with a standard scaler across the participants for HMM training.

As shown in \figurename~\ref{fig:HMM_result}, the HMM consistently identified four latent behavioral states. State \textit{S3} exhibited the strongest unhealthy dietary signature, with substantially higher soda consumption ($0.60$ items/day compared to $\leq 0.20$ items/day in the remaining states). State \textit{S2} was characterized by the highest level of physical activity among overweight and obese participants ($231.3$ activity events/day versus $\leq 49.9$ in the other states), whereas activity intensity was comparatively lower for the normal and underweight groups. State \textit{S1} represented a largely inactive state, with most behavioral measures remaining close to zero. Finally, State \textit{S0} was associated with high digital engagement accompanied by heavy eating among overweight and obese individuals, but healthy eating among normal and underweight individuals. Based on this, we interpret \textit{S0} as active eating, \textit{S1} as low engagement/inactive, \textit{S2} as exercise/digital exposure, and \textit{S3} as unhealthy routine/drinking. These labels are descriptive interpretations assigned post hoc based on the estimated emission means, rather than predefined behavioral categories.

As shown in \figurename~\ref{fig:normal_transition}, normal and underweight individuals predominantly occupy the low engagement/inactive state \textit{S1} (occupancy $63\%$, recurrence $84.68\%$). Importantly, when they entered the active/healthy eating \textit{S0}, this state remained moderately stable (recurrence $40.54\%$). These individuals, when preferred ($8\%$ from \textit{S0}), consistently engage in periodic low-intensity exercise \textit{S2} (recurrence $35.59\%$). Transitions directly from the healthy routine to the Unhealthy state \textit{S3} are very small, as exercise days primarily transitioned back to either the inactive state ($31\%$) or the healthy eating state ($24\%$), with only $7\%$ progressing toward the unhealthy behavioral state. \textit{Thus, normal and underweight individuals lead a sustainable lifestyle in which they eat healthy, exercise occasionally, and avoid junk food}.

In contrast, overweight and obese individuals exhibited a different behavioral trajectory as shown in \figurename~\ref{fig:obesity_transition}. When they entered the active/heavy eating state \textit{S0}, this behavior continues (recurrence $47.46\%$), indicating prolonged engagement in heavy eating and food-related digital activities. Although individuals occasionally transitioned from \textit{S0} to the exercise/digital exposure state \textit{S2} ($5\%$), this exercise state was short-lived (recurrence $25.11\%$). Rather than sustaining physical activity or returning to healthier eating habits, participants frequently relapsed from \textit{S2} back to the heavy-eating state \textit{S0} ($24\%$) or transitioned to the unhealthy state \textit{S3} ($22\%$), which is characterized by high soda consumption. Furthermore, the unhealthy state often reverted to heavy eating ($18\%$), reinforcing this vicious cycle. Unlike normal-weight individuals, who typically return to healthy eating after exercise, \textit{overweight and obese individuals exhibit a recurring pattern of heavy eating, short-lived exercise, and relapse into unhealthy dietary behaviors}. This self-reinforcing behavioral dark pattern makes sustained lifestyle change and long-term weight loss particularly challenging.

\section{Limitations and Future Work}

Our analysis should be interpreted as purely observational, based on a cross-country behavioral dataset of individuals with normal and high BMI. We discuss several limitations of this study as follows:
\begin{itemize}
\item BMI distribution of \textit{DiversityOne} dataset is highly imbalanced, with 42 underweight, 326 normal-weight, 70 overweight, and 15 obese (i.e., 368 normal and 85 high BMI) individuals. This data imbalance limits the statistical power of our analysis for obesity-specific behavioral patterns.
\item The dataset consists exclusively of university students, whose daily routines, living environments, and smartphone usage patterns may differ substantially from those of the broader adult population. Therefore, the findings may not generalize to children or older adult populations.
\item The current analysis overlooks dietary habits and digital behaviors that are significantly influenced by local food environments, cultural practices, and the availability of food-delivery services in specific countries. 
\end{itemize}

As a future direction, our findings should be supported by larger, more diverse cohorts sampled from general adult populations across several countries, enabling country-stratified analysis. Moreover, exploring additional contextual information, such as mood and mental well-being, may further improve our understanding of the behavioral factors associated with body weight.

\section{Conclusion}
In this work, we analyzed the \textit{DiversityOne} dataset, comprising four weeks of passive smartphone sensing and ecological momentary assessments from $453$ participants across eight countries, to investigate the relationship between everyday behaviors and BMI. Our findings show that dietary choices, particularly the consumption of soda, alcohol, and processed meat, together with increased use of food-delivery apps, exhibit the strongest associations with higher BMI. Further, our Hidden Markov Model uncovered distinct behavioral routines across BMI groups. While normal-weight individuals generally follow a sustainable cycle of healthy eating and periodic exercise, overweight and obese individuals exhibit a recurring behavioral dark pattern in which short-lived exercise is frequently followed by relapse into unhealthy eating and drinking behaviors, making weight loss challenging.

\section*{Acknowledgement}
The authors would like to thank the anonymous reviewers for the constructive comments, which have helped to improve the overall presentation of the paper. We also thank the authors of the \textit{DiversityOne} dataset for making this study possible.

\balance

\bibliographystyle{ACM-Reference-Format}
\bibliography{references.bib}
\end{document}